\documentclass{PoS}
\usepackage{setspace}
\usepackage{graphicx}
\usepackage{dcolumn}
\usepackage{multirow}
\usepackage{subfigure}
\usepackage{times,mathptm}
\usepackage{float}
\usepackage{color}
\usepackage{amsmath,amsfonts}
\usepackage{mathptmx}
\usepackage{mathrsfs}
\usepackage{bbm}
\usepackage{bm}
\usepackage{xfrac}

\newcommand{\bea}{\begin{eqnarray}}
\newcommand{\eea}{\end{eqnarray}}

\renewcommand{\d}{\delta}
\newcommand{\ihat}{\boldsymbol{\hat{\textbf{\i}}}}
\renewcommand{\l}{\lambda}

\newcommand{\tQ}{\widetilde{Q}}

\renewcommand{\b}{\beta}
\renewcommand{\a}{\alpha}

\newcommand{\tr}{\text{Tr}}

\newcommand{\tP}{\widetilde{P}}

\newcommand{\vx}{{\vec{x}}}
\newcommand{\vy}{{\vec{y}}}

\newcommand{\vk}{{\vec{k}}}

\newcommand{\m}{\mu}

\newcommand{\s}{\sigma}
\renewcommand{\k}{\kappa}

\newcommand{\D}{\Delta}

\newcommand{\vq}{\vec{q}}

\newcommand{\N}{{\cal N}}

\newcommand{\oh}{\frac{1}{2}}

\newcommand{\dg}{\dagger}
\newcommand{\non}{\nonumber}

\newcommand{\rf}[1]{(\ref{#1})}
\newcommand{\ra}{\rightarrow}
\newcommand{\pa}{\partial}
\renewcommand{\vec}[1]{\bm #1}
\usepackage{ulem}
\title{Extracting the effective Polyakov line action from SU(2) and SU(3) lattice gauge theories}

\ShortTitle{Effective Polyakov line action}

\author{\speaker{Jeff Greensite}%
         \thanks{Supported in part by 
the U.S.\ Department of Energy under Grant No.\ DE-FG03-92ER40711.}\\
        Physics and Astronomy Dept., San Francisco State
University, San Francisco, CA~94132, USA\\
        E-mail: \email{greensit@sfsu.edu}}

\author{Kurt Langfeld \thanks{Supported by STFC under the DiRAC framework and by the HPCC Plymouth, where the numerical computations have been carried out.}\\
        School of Computing \& Mathematics, University of Plymouth, Plymouth, PL4 8AA, UK\\
        E-mail: \email{kurt.langfeld@plymouth.ac.uk}}

\abstract{We describe the ``relative weights'' method used to compute the effective Polyakov line action corresponding
to a given lattice gauge theory, and present some results that have been obtained so far.  The main motivation is the sign problem, which may be easier to address in the effective theory than in the underlying gauge theory.}

\FullConference{QCD-TNT-III-From quarks and gluons to hadronic matter: A bridge too far?,\\
		2-6 September, 2013\\
		European Centre for Theoretical Studies in Nuclear Physics and Related Areas (ECT*), Villazzano, Trento (Italy)}

\begin{document}

\section{The effective Polyakov line action}

   The effective Polyakov line action (PLA) $S_P$ is obtained from the underlying lattice gauge theory by integrating out all degrees of freedom subject to the constraint that the Polyakov line holonomies are held fixed.  In temporal gauge we have 
\bea
\exp\Bigl[S_P[U_{\vx}]\Bigl] =    \int  DU_0(\vx,0) DU_k  D\phi ~ \left\{\prod_{\vx} \d[U_{\vx}-U_0(\vx,0)]  \right\}
 e^{S_L} \ ,
\label{S_P}
\eea
where $\phi$ denotes any matter fields, scalar or fermionic, coupled to the gauge field, and $S_L$ is the lattice action.  Our interest in the PLA is due to its possible application to the sign problem.  Using the strong-coupling/hopping parameter expansion, one finds at lowest order that $S_P$ has the form
\bea
S_P =   \b_P \sum_{\vx} \sum_{i=1}^3  [\tr U_\vx^\dg \tr U_{\vx+\ihat} + \tr U_\vx  \tr U^\dg_{\vx+\ihat}] 
   + \k \sum_\vx [e^{N_t\m} \tr U_\vx + e^{-N_t\m} \tr U^\dg_\vx]  \ ,
\label{strong}
\eea
where $\b_P$ and $\k$ can be expressed in powers of $\b$ and the hopping parameter $h$.  An action of this form, disregarding its origin, seems
to have a relatively mild sign problem, for a large range of parameters $\b_P,\k,\m$, and has been solved by various means, including dual representations \cite{Mercado:2012ue}, stochastic quantization \cite{Aarts:2011zn}, reweighting \cite{Fromm:2011qi}, and mean field methods \cite{Greensite:2012xv}.  The problem we will address is how to derive
the PLA corresponding to a given lattice gauge theory when the lattice coupling is not so strong, and the hopping parameter
is not small.  It is actually only necessary to derive the PLA at chemical potential $\m=0$, because once the PLA at $\m=0$ is known, the PLA at non-zero $\m$ is obtained from a simple substitution 
 \bea
     S_P^\m[U_\vx,U^\dg_\vx] =  S_P^{\m=0}[e^{N_t \m} U_\vx,e^{-N_t \m}U^\dg_\vx]   \ .
\label{convert}
\eea
One can show \cite{Greensite:2013yd} that this relationship is true to all orders in the strong-coupling/hopping parameter computation of $S_P$, and we will assume that it holds in general.  The method we use to derive the PLA at $\m=0$, to be expained below, we call ``relative weights."  This talk is based on work reported in refs.\ \cite{Greensite:2013yd,Greensite:2013bya} and, for the SU(3) group, on work in progress.

    There have been other approaches to calculating the effective Polyakov line action, including
strong-coupling expansions \cite{Fromm:2011qi,Bergner:2013rea}, the Inverse Monte Carlo method \cite{Heinzl:2005xv,Wozar:2007tz},
and the Demon approach \cite{Velytsky:2008bh,Wozar:2008nv}, resulting in effective actions of varying complexity.
A crucial test of any approach is to calculate the Polyakov line correlator
\bea
G(R) = \langle P_\vx  P_\vy  \rangle ~~ \text{with} ~~R=|\vx-\vy| ~~~\text{and}~~~P_\vx = {1\over N} \tr[U_\vx]
\eea
in both the effective action and the underlying gauge theory, and see if these agree.  We do not believe that accurate agreement has been demonstrated in these approaches at the larger $\b$ values, at least not beyond separations $R$ of two or three lattice spacings.  

\section{Relative Weights}

       Let $U_\vx$ at all $\vx$ on the $D=3$ dimensional lattice represent a configuration of Polyakov line holonomies, and consider
any path through the space of all such configurations $U_\vx(\l)$ parametrized by $\l$.  The relative weights method allows us to compute the derivative $dS_P/d\l$ along the path, and from such derivatives we try to deduce the PLA
$S_P$ itself. 

    Let $U'_\vx, U''_\vx$ denote two configurations along the the path, corresponding to $\l+\oh \D \l$ and $\l-\oh \D \l$
respectively.   We define the action difference $\D S_P = S_P[U'_\vx] - S_P[U''_\vx]$, and also lattice actions $S_L$ in temporal gauge with fixed holonomies
\bea
S_L[U'] \equiv S_L\Bigl[U_0(\vx,0)=U'_\vx \Bigr] ~~~,~~~
S_L[U''] \equiv S_L\Bigl[U_0(\vx,0)=U''_\vx \Bigr] \ ,
\eea
i.e.\ the timelike link variables on the $t=0$ timeslice are held fixed to either $U'_\vx$ or  $U''_\vx$; these links are not integrated over in the path integration.  Then, from eq.\ \rf{S_P} we have
\bea
e^{\D S_P} &=&  {\int  DU_k  D\phi ~  e^{S'_L} \over \int  DU_k  D\phi ~  e^{S''_L} } 
={\int  DU_k  D\phi ~  \exp[S'_L-S''_L] e^{S''_L} \over \int  DU_k  D\phi ~  e^{S''_L} }
\non \\
&=& \Bigl\langle  \exp[S'_L-S''_L] \Bigr\rangle'' \ ,
\eea
where the notation $\langle...\rangle''$ indicates that the expectation value is evaluated in the measure
proportional to $e^{S''_L}$.  We then have
\bea
\left({dS_P \over d \l}\right)_{\l=\l_0} \approx {\D S_P \over \D \l} \ .
\eea
The question is which path derivatives will help us to determine $S_P$ itself.

   Let us start with the gauge group SU(2).  There is no sign problem in this case, but our aim is right now is to see if we can extract the PLA by the method described.  The SU(2) PLA can only depend on Polyakov lines $P_\vx = \oh \tr U_\vx$.  Make a Fourier expansion
\bea
                P_{\vx} =  a_0 + \oh \sum_{\vq\ne 0} \Bigl\{ a_{\vq} \cos(\vq \cdot \vx) + 
                       b_{\vq} \sin(\vq \cdot \vx) \Bigr\}  \ .
\eea                       
Then we compute $(\pa S_P/\pa a_\vk)_{a_\vk=\a}$ by the relative weights method at a ``typical'' point in configuration space, i.e.\ a thermalized configuration generated by lattice Monte Carlo, by the following procedure:  (1) generate a thermalized lattice configuration $U_\m(x)$ by the usual methods, and set $U_\vx = U_0(\vx,0)$.  (2) Fourier decompose
$P_\vx$ and set $a_\vk=0$ for some given $\vk$.  Call the resulting configuration, transformed back to position space,
$\widetilde{P}_\vx$.  Then construct
\bea
P'_\vx &= (\a + \oh \D a_\vk)\cos(\vk \cdot \vx) + f \tP_\vx
\non \\
P''_\vx &= (\a - \oh \D a_\vk)\cos(\vk \cdot \vx) + f \tP_\vx \ ,
\eea
where $f=1-\a$.  (4)  Derive, from the Polyakov line configurations $\tP'_\vx$ and $\tP''_\vx$ the corresponding Polyakov line holonomies $U'_\vx$ and $U''_\vx$. (5) Compute  $(\pa S_P/\pa a_\vk)_{a_\vk=\a} \approx \D S_P/\D a_\vk$ by the relative weights technique described above. 

\begin{figure}[ht]
\centering
\subfigure[~ $\a=0.05$ with linear fit]{
\resizebox{71mm}{!}{\includegraphics{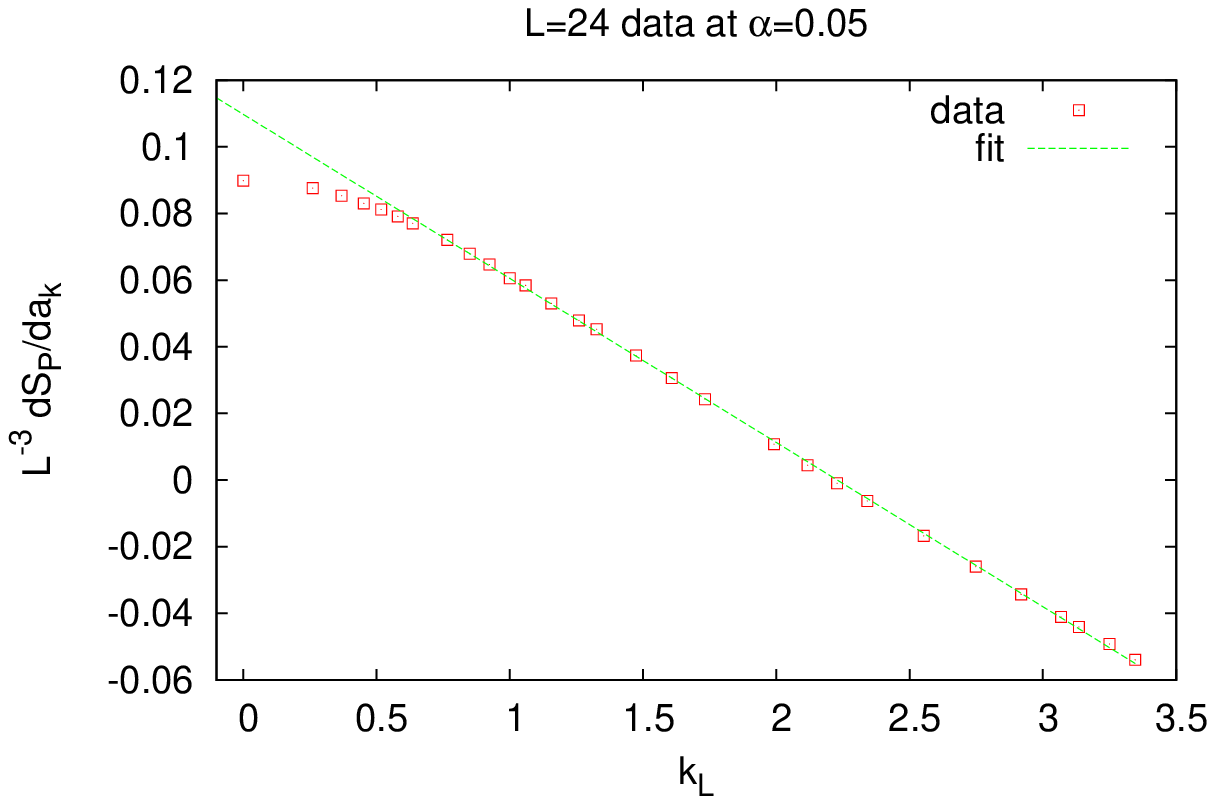}}
\label{d24a05}
}
\subfigure[~ $\a$-scaling]{
\resizebox{71mm}{!}{\includegraphics{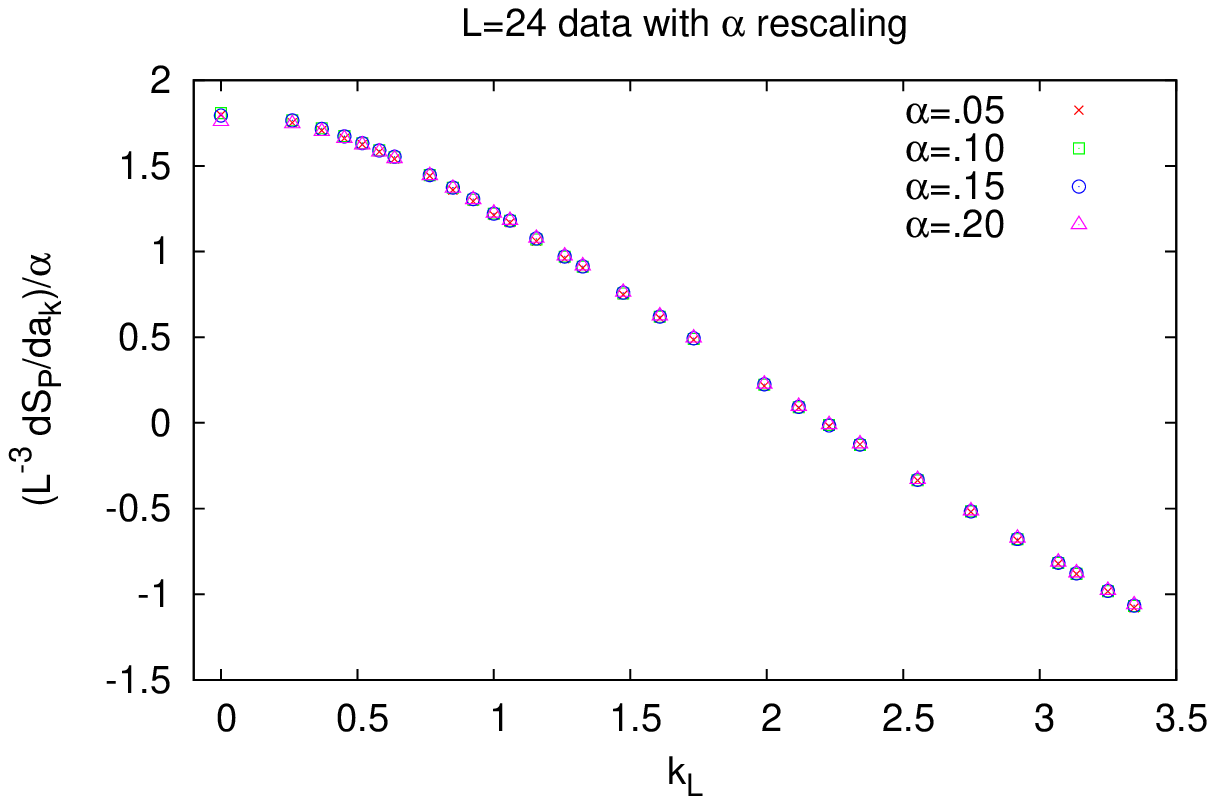}}
\label{scaling}
}
\caption{(a) Derivatives of the PLA $L^{-3} \pa S_P/\pa a_{\vk}$ evaluated at $a_{\vk}=\a=0.05$, vs.\ lattice momenta $k_L$.  Also shown is a linear best fit to the data at $k_L > 0.7$. (b) Derivatives $L^{-3} (\pa S_P/\pa a_{\vk})_\a$ divided by $\a$, vs.\ lattice momenta $k_L$, for $\a=0.05,0.10,0.15,0.20$.  It is clear that the derivatives of $S_P$ depend linearly on $\a$.} 
\label{d24}
\end{figure} 

\section{SU(2) pure gauge theory}

\begin{figure}[t]
\centerline{\scalebox{0.8}{\includegraphics{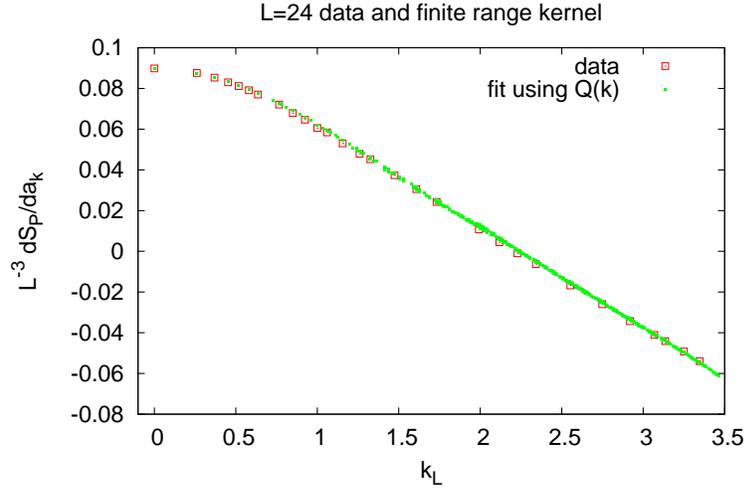}}}
\caption{A test of eq.\  (3.2) at $\a=0.05$.  The derivative data of Fig.\ 1 is plotted against
the conjectured fitting function $\a(\oh c_1 - 2c_2 \tQ(k_L))$ with $r_{max}=3$ }
\label{kernel}
\end{figure}

    We begin with pure SU(2) gauge theory at $\b=2.2$ on a $24^3 \times 4$ lattice volume.  At this extension $N_t=4$ in the time direction, the deconfinement transition is very close to $\b=2.3$.   Figure \ref{d24a05} shows our data obtained on this lattice 
for the path derivative $L^{-3}(\pa S_P/\pa a_\vk)_{a_\vk=\a}$, evaluated at $\a=0.05$, versus the lattice momentum $k_L$, defined from wavenumbers $\vec{k}$ as  $k_L = 2\sqrt{\sum_{i=1}^3 \sin^2(\oh k_i)}$.  Here $L^3=24^3$ is the volume of a time slice. What is striking about this data is that apart from low momenta, the data fits very accurately onto a straight line.  
Figure \ref{scaling} is the same observable on the $y$-axis divided by $\a$, for several different values of $\a$.  From the fact that the data points at each $\a$ coincide, it is clear that the derivative must be linear in $\a$, which means that $S_P$ itself is quadratic in each momentum mode.  It follows that $S_P$ is bilinear in the Polyakov lines, and can be written in the form
\bea
            S_P =  \oh c_1 \sum_{\vx} P^2_{\vx} -  2c_2 \sum_{\vx \vy} P_{\vx} Q(\vx - \vy) P_{\vy}  \ .
\label{PLA}
\eea
Let  $\tQ(\vk)$ be the finite Fourier transform of the kernel $Q$.  We find that $\tQ(\vk)$ depends only on the magnitude $k_L$,
and that for a PLA of the form \rf{PLA}\footnote{The relative factor of two between $k_L=0$ and $k_L >0$ is reflects the fact
that $\sum_x \cos^2(\vk \cdot \vx) = \oh L^3$ while $\sum_x 1 = L^3$. The data points appearing on the plots at $k_L=0$ is the
data value divided by two.}
\bea
{1\over L^3} \left({d S_P[U_\vx(a_{\vk})] \over da_{\vk}}\right)_{a_{\vk}=\a} &=&  \left\{
   \begin{array}{cc}
      \a(\oh c_1 - 2c_2 \tQ(k_L)) & k_L \ne 0 \cr
       & \cr
      2\a(\oh c_1 - 2c_2 \tQ(0))   &   k_L=0 \end{array} \right. \ .
\eea
From Fig.\ \ref{d24} we see that $\tQ(k_L) \sim k_L$ except at small $k_L$.  If it were true that $\tQ(k_L)=k_L$ at all $k_L$,
we would have $Q(\vx-\vy)=\left(\sqrt{-\nabla^2_L}\right)_{\vx \vy}$, where $\nabla^2_L$ is the lattice Laplacian.  But then the
kernel $Q(\vx-\vy)$ would be long-range, which would violate one of the assumptions of the Svetitsky-Yaffe conjecture
\cite{Svetitsky:1985ye}, and in
any case we see that the data deviates from linearity at small $k_L$.  So we implement a finite range condition in the simplest way, choosing
\bea
          Q(\vx-\vy) = \left\{  \begin{array}{cc}
                     \Bigl(\sqrt{-\nabla_L^2}\Bigr)_{\vx \vy}  &  |\vx-\vy| \le r_{max} \cr
                       0 & |\vx-\vy| > r_{max} \end{array} \right. \ .
\label{Q}
\eea
Then we Fourier transform to obtain $\tQ(k_L)$, and select the value of $r_{max}$ which best fits the data.  The constants
$c_1, c_2$ are determined from the straight-line fit through the higher momentum data.  At $\b=2.2$ and $N_t=4$, the constants $c_1=4.417,c_2=0.498$ and $r_{max}=3$ give an excellent fit to the data as seen in Fig.\ \ref{kernel}.

\begin{figure}[ht]
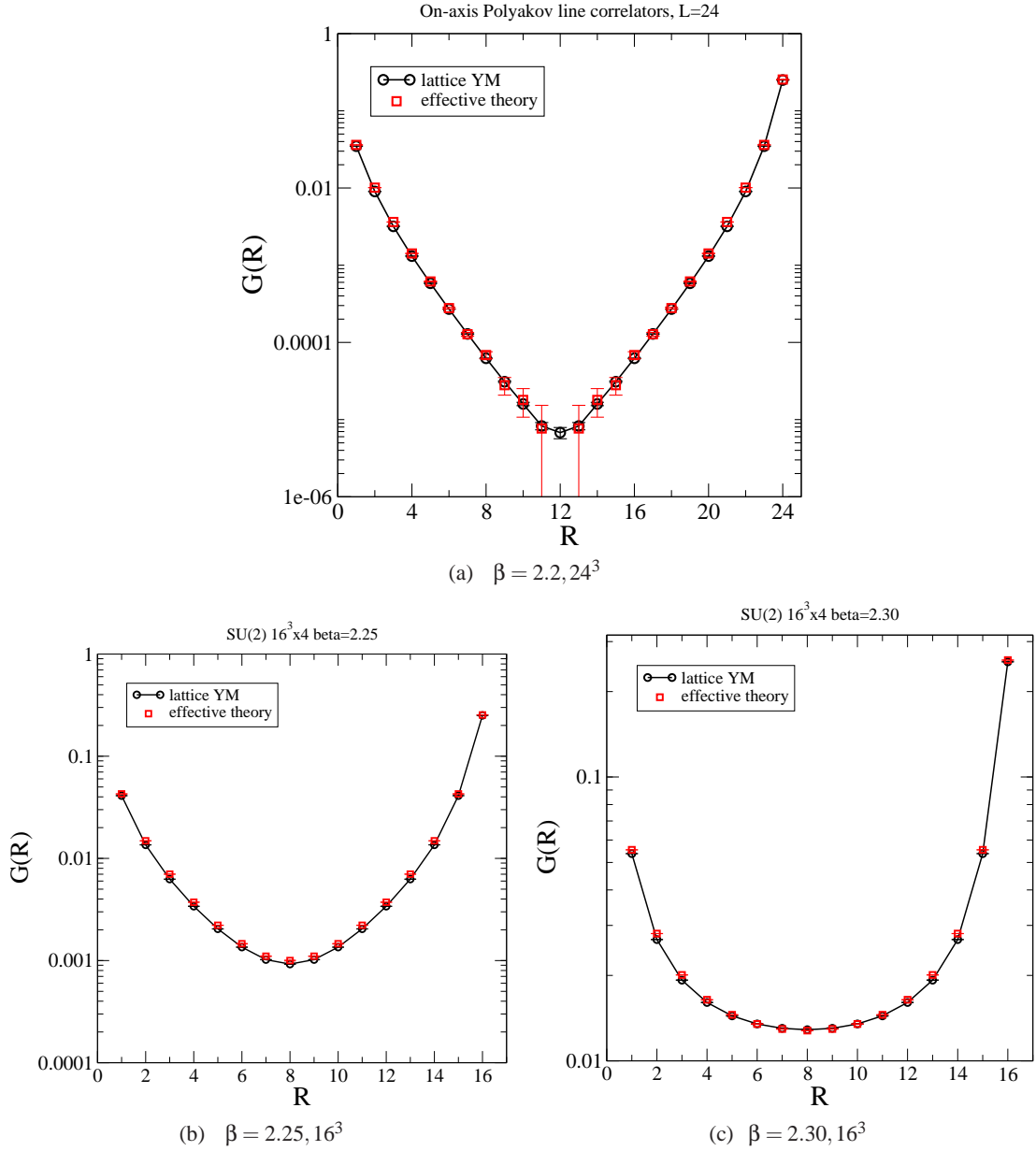

\centering
\subfigure[~ $\b=2.2, 24^3$]{
\resizebox{80mm}{!}{\includegraphics{corrax24.eps}}
\label{corr220}
}
\subfigure[~ $\b=2.25, 16^3$]{
\resizebox{71mm}{!}{\includegraphics{pot16_225.eps}}
\label{corr225}
}
\subfigure[~$\b=2.30, 16^3$]{
\resizebox{71mm}{!}{\includegraphics{pot16_230.eps}}
\label{corr230}
}
\caption{Polyakov line correlators derived from the underlying lattice gauge theory (black circles) on an $L^3 \times 4$ lattice, and from the effective PLA (red squares) on an $L^3$ lattice. (a) $\b=2.2$ and $L=24$. (b) $\b=2.25$ and $L=16$. (c) $\b=2.3$ and $L=16$. This coupling is at the deconfinement transition.} 
\label{corr}
\end{figure}

    Given $c_1,c_2,r_{max}$ the effective PLA is determined, and the crucial question is whether Polyakov line correlators obtained in the effective theory agree with the same correlators determined in the underlying lattice gauge theory.  In Fig.\ \ref{corr} we show our results for $N_t=4$ lattice spacings in the time direction at $\b=2.2,2.25,2.3$.  The last coupling is right at the deconfinement transition.  It can be seen that agreement between the Polyakov line correlators is very accurate, with agreement down to $O(10^{-5})$.
    
   The appearance of $\sqrt{-\nabla^2_L}$ in the kernel $Q(\vx-\vy)$ is striking, and has not been clearly seen in other approaches  \cite{Fromm:2011qi,Heinzl:2005xv,Wozar:2007tz,Velytsky:2008bh,Wozar:2008nv} to extracting the effective PLA.  It is worth asking if this behavior of the kernel should be expected for some reason, at least for small separations.  To at least partially answer this question, let us consider a much simpler field theory, namely a massless scalar free field theory.  Motivated by the definition of the effective PLA in \rf{S_P}, which involves integrating out all degrees of freedom apart from timelike links at $t=0$, let us consider the analogous exercise of
integrating out all degrees of freedom in the scalar free field theory, except for those at time $t=0$.  It is well known that the result is simply the square of the ground state wavefunctional
\bea
\Psi_0^2[\phi_\vx] &=& \int D\phi ~ \prod_x \d[\phi(\vx,0) -\phi_\vx] 
\non \\
& & \qquad \times \exp\left[- \oh \int d^3x dt \phi(\vx,t) (-\pa^2) \phi(\vx,t) \right] \ .
\eea
The functional integral over $\phi(\vx,t\ne 0)$ can be carried out analytically, with the result
\bea
\Psi_0^2[\phi_\vx] = \N \exp\left[- \int d^3x d^3y ~ \phi_\vx \left(\sqrt{-\nabla^2}\right)_{\vx \vy} \phi_\vy \right] \ ,
\eea
where $\N$ is a normalization constant.  Note the appearance of the non-local kernel $\sqrt{-\nabla^2}$.  In an asymptotically free gauge theory we might also expect to see, at weak couplings, the kernel $\sqrt{-\nabla^2}$ in the PLA at small separations.

\begin{figure}[t]
\centerline{\scalebox{0.6}{\includegraphics{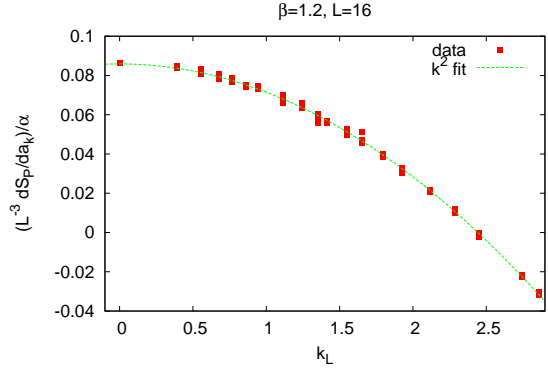}}}
\caption{Comparison of the best fit $c_1/2 - 2c_2 k_L^2$ to the relative weights data at a strong coupling $\b=1.2$.}  
\label{comfit}
\end{figure}

   As a further check of our methods we can also compute the PLA at small $\b$, where the effective PLA, of the form 
\rf{strong}, can be computed from the lattice strong coupling expansion.  Our $\pa S_P/\pa a_\vk$ data for $\b=1.2$ is 
shown in Fig.\ \ref{comfit}. In this case the data fits a parabola, $\oh c_1 - 2c_2 k_L^2$, rather than a straight line, which implies that $G(\vx-\vy) = (-\nabla^2)_{\vx \vy}$, and this is a nearest-neighbor coupling, as in \rf{strong}.  The comparison of the PLA extracted from this data to the PLA derived from a strong-coupling expansion shows very good agreement:
\bea
S_P = \left\{ \begin{array}{ll}
            0.02859(3)  \sum_\vx \sum_{i=1}^3   P_\vx P_{\vx+ \ihat} & \mbox{relative weights} \cr \cr
            0.02850  \sum_\vx \sum_{i=1}^3   P_\vx P_{\vx+ \ihat} & \mbox{strong coupling} \end{array} \right. (\b=1.2) \ .
\eea

\section{Adding a matter field}

\begin{figure}[ht]
\centering
\subfigure[~ full range]{
\resizebox{65mm}{!}{\includegraphics{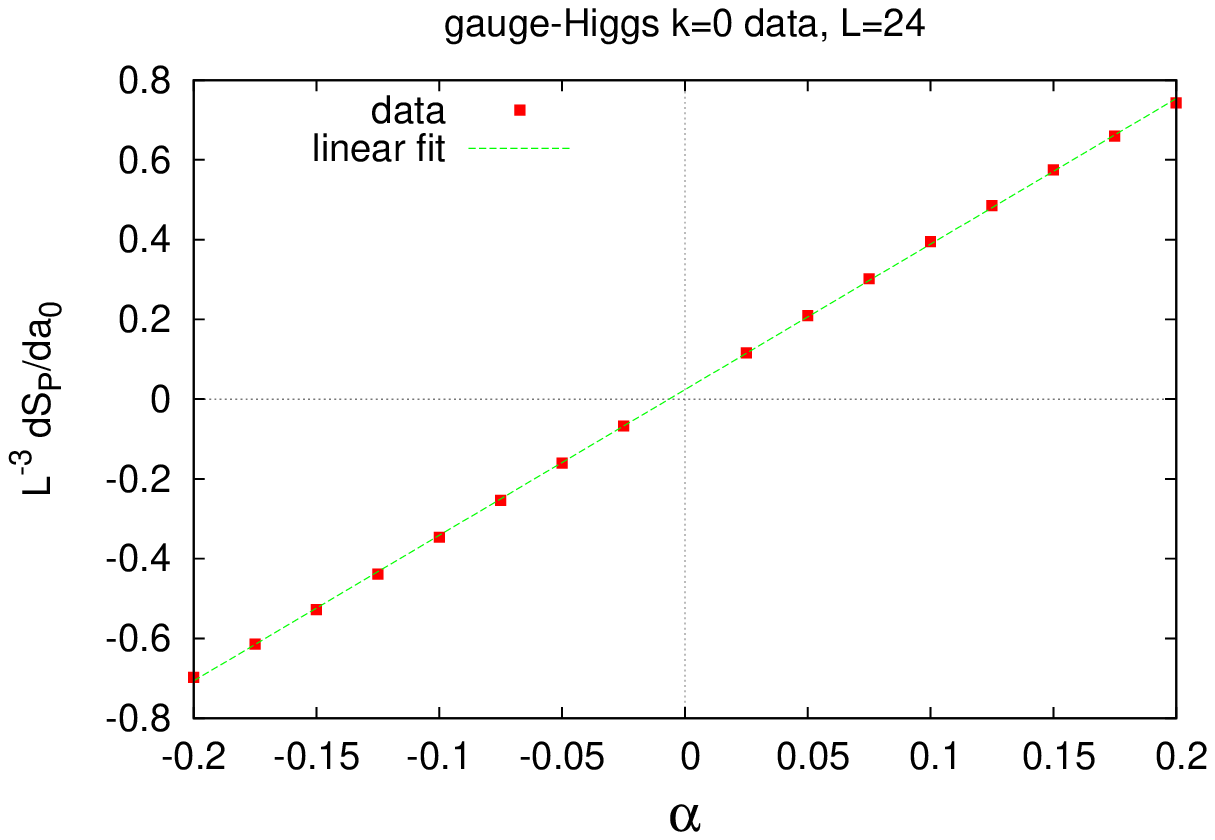}}
\label{zmodeh1}
}
\subfigure[~ close-up]{
\resizebox{65mm}{!}{\includegraphics{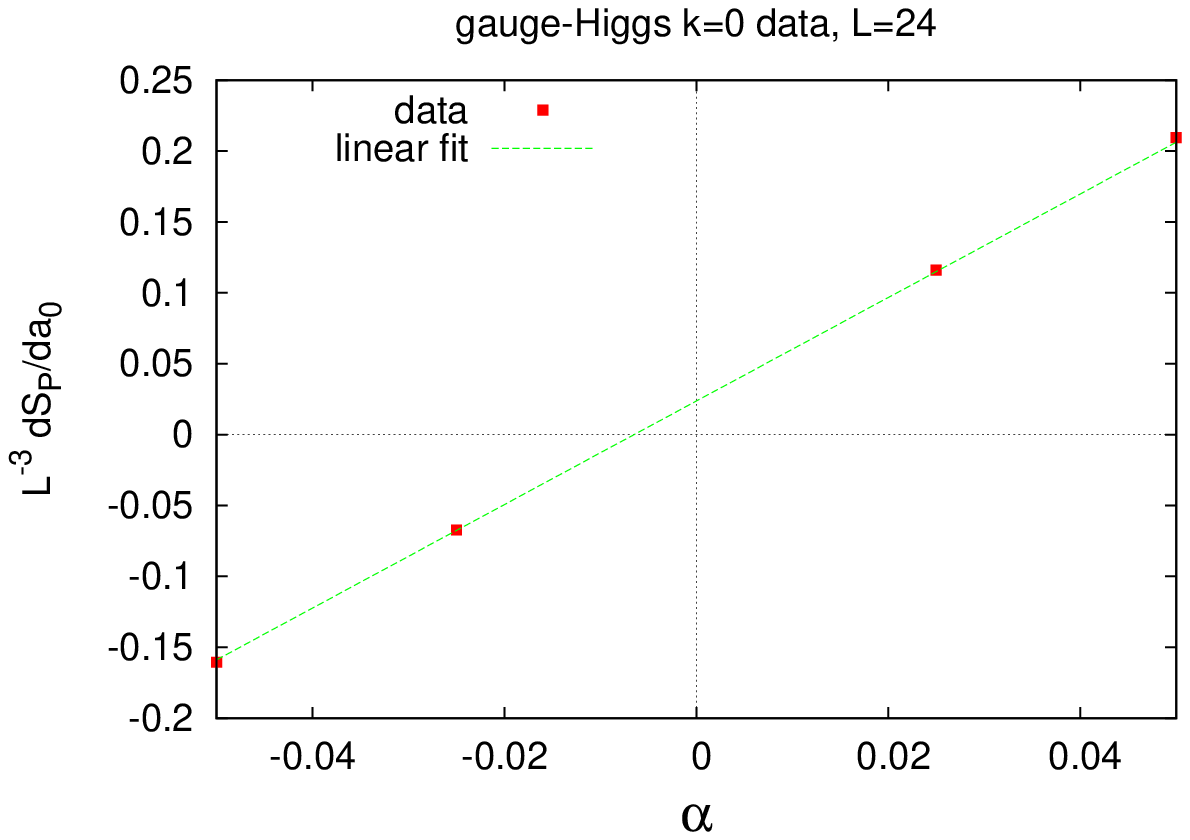}}
\label{zmodeh2}
}
\caption{The derivatives of $S_P$ with respect to the amplitude of the zero mode in the gauge-Higgs theory, evaluated at positive and negative values of $a_0=\a$.  (a) shows the full range of the data; (b) is a closeup near $\a=0$.  The $y$-intercept of this data is non-zero, and determines the coefficient $c_0$ of the linear, $Z_2$-symmetry breaking term in the effective PLA (3.2).}
\label{zmodeh}
\end{figure}

   We now add a fixed modulus Higgs field in the fundamental representation of SU(2), which breaks the global center symmetry.
For an SU(2) gauge group, the corresponding gauge-Higgs action can written
\bea
    S_L = \b \sum_{plaq} \oh \mbox{Tr}[UUU^{\dg}U^{\dg}] + \k \sum_{x,\m} \oh
              \mbox{Tr}[\phi^\dg(x) U_\m(x) \phi(x+\widehat{\m})] \ ,
\label{gh_action}
\eea
and we work at $\k=0.75$ and $\b=2.2$ on a $24^3 \times 4$ lattice.  This time the PLA picks up a center symmetry-breaking term which is linear in the Polyakov line variable
\bea
           S_P =   c_0 \sum_{\vx} P_{\vx} + \oh c_1 \sum_{\vx} P^2_{\vx} -  2c_2 \sum_{\vx \vy} P_{\vx} Q(\vx - \vy) P_{\vy} \ .
\label{SP2}
\eea
In the Fourier decomposition, the symmetry-breaking term is linear in $a_0$, and it implies that $\pa S_P/\pa a_0$, evaluated at $a_0=\a$, goes to a non-zero constant in the $\a \ra 0$ limit.  The coupling $c_0$ is given by the extrapolation of the 
$L^{-3}(\pa S_P/\pa a_0)$ data to $\a=0$, as shown in Fig.\ \ref{zmodeh}.  The center symmetry-breaking term does not contribute
at $k_L \ne 0$, and $c_1,c_2,r_{max}$ are determined as in the pure gauge case.

\begin{figure}[h!]
\centerline{\scalebox{0.4}{\includegraphics{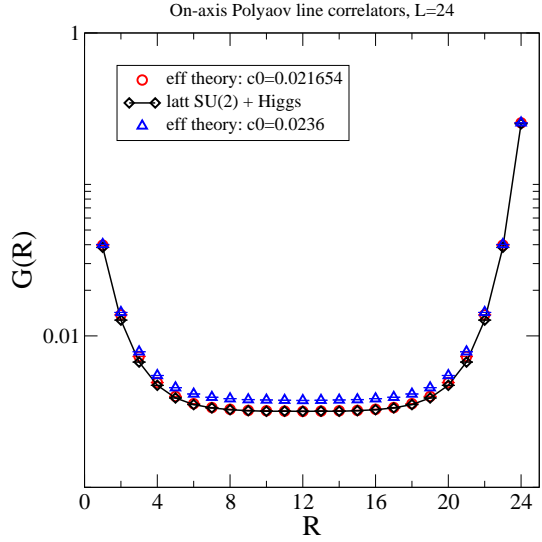}}}
\caption{A comparison of the Polyakov line correlation functions $G(|\vx-\vy|) = \langle P_{\vx}  {P_\vy} \rangle$ as computed
via lattice Monte Carlo simulation of the underlying gauge-Higgs theory (black diamonds) on a $24^3 \times 4$ lattice, 
at couplings $\b=2.2,~\k=0.75$, and via Monte Carlo simulation of the corresponding effective action $S_P$ of eq.\ (3.2)
(blue triangles, $c_0=0.0236$).  Also shown is a simulation of the effective action with a slightly different value of $c_0=.02165$ (red circles).}
\label{corrH24}
\end{figure}  

\begin{figure}[h!]
\centerline{\scalebox{0.8}{\includegraphics{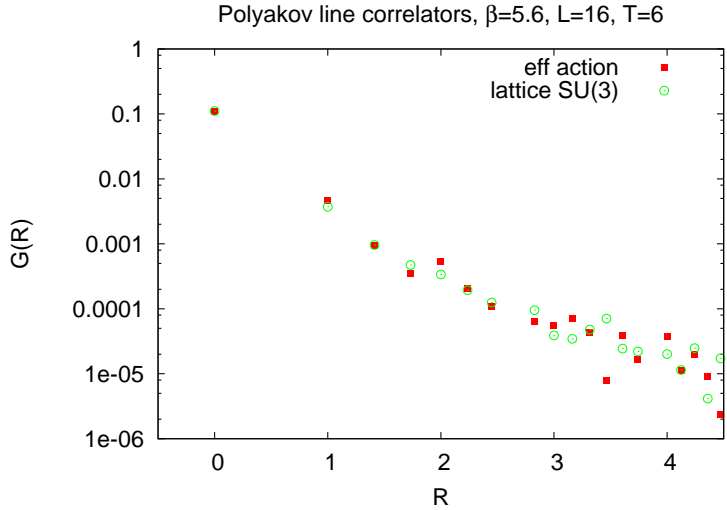}}}
\caption{A comparison of the off-axis SU(3) Polyakov line correlators computed in the effective PLA (solid circles), and in the underlying lattice SU(3) pure gauge theory at $\b=5.6$ on a $16^3 \times 6$ lattice (open circles).}
\label{corr56T6}
\end{figure}  
   
    Our result for the Polyakov correlator (blue triangles), compared to the corresponding correlator in the underlying lattice gauge theory (black circles) is shown in Fig.\ \ref{corrH24}.  Agreement is quite good, using our value of $c_0$ determined to be
$c_0=0.0236(14)$.  We can get near perfect agreement with the underlying lattice gauge theory correlator by setting $c_0=0.0265$ (red circles), which is about $1.4 \s$ away from our calculated value.

\section{Next Steps}

    There is no sign problem in SU(2) gauge theory with matter fields.  This is due to the pseudo-real property of SU(2) group respresentations.  Our focus here on SU(2) is for testing purposes:  we want to check if the relative weights method can be used to extract the corresponding effective Polyakov line action.  All indications suggest that method can indeed be used for that purpose.
    
     The next step is to move on to SU(3) gauge theory which, if the gauge field is coupled to matter fields with non-vanishing
N-ality, will have a sign problem at finite chemical potential.  Here again the first task it to extract the PLA for the pure gauge theory.  A very preliminary result is shown in Fig.\ \ref{corr56T6}.  This is a comparison of off-axis Polyakov line correlators in the PLA and in the underlying lattice gauge theory at $\b=5.6$ and lattice volume $16^3 \times 6$, where the PLA has been determined by the same methods used in the SU(2) case.   It is desirable to try out other values of $\b$, and then add in matter fields.  First we would introduce a scalar field in the fundamental representation, as in the SU(2) case, and if this works out we would move on to
fermions.  All of the simulations are done at $\m=0$, but we stress again that the $\m\ne 0$ PLA is obtained from $\mu=0$ by
the simple substitution \rf{convert}.  The final step in the program, if it works up to this point, would be to obtain the phase diagram of the SU(3) theory in the $\m-T$ plane, by simulating the PLA by any of the methods  \cite{Mercado:2012ue,Aarts:2011zn,Fromm:2011qi,Greensite:2012xv} that have been applied successfully to the nearest-neighbor form of the Polyakov line action at finite chemical potential.

\bibliography{pline}

\end{document}